\documentclass[aps,amsmath,twocolumn,prl]{revtex4}
\usepackage{graphics, setspace,epsfig,color,subfigure}
\usepackage[letterpaper, dvips,width=7.5in,height=8.5in,includemp=false]{geometry}
\usepackage[vcentermath]{}

\newcommand{\be}{\begin{equation}}
\newcommand{\ee}{\end{equation}}
\begin{document}
{\large \bf \noindent Response to D.T.~Son's comment on ``Is there
a `most perfect fluid' consistent with quantum field theory?''
\\}

D.T.~Son raises an extremely interesting and subtle point in his
comment[1]. However, the conclusion in the original letter that
theoretically consistent exceptions exist for the proposed general
bound that $\eta/s \ge (4 \pi)^{-1}$ for all fluids [2], appears
to remain unaffected by the issue raised.

Firstly, in the context of nonrelativistic quantum mechanics there
exist theoretically consistent systems  which violate the putative
bound [3] and are unaffected by the issue raised in the comment.
Thus, the origin of any general bound must lie beyond quantum
mechanics. One suggestion is that relativistic quantum field
theories with sensible behavior in the ultraviolet somehow act as
a censor by preventing {\it any} nonrelativistic system arising
from such a field theory to violate the bound. It was pointed out
in ref. [3] that  the ultraviolet behavior of many-flavored QCD
acts this way in preventing a pion gas from violating the bound.
However, it appears unlikely  {\it a priori} that the bound is
fundamentally tied to relativistic field theoretic effects, if for
no other reason than the fact, first noted in Ref. [2], that $c$
is absent from the bound.  Thus, there are deep reasons to doubt a
general bound regardless of the issue raised in [1].

It was suggested in Ref.~[3] that if a bound on $\eta/s$ is
general, it is natural to expect it to hold (up to possible small
violations due to ambiguities) for metastable fluids provided that
the fluid is in a sufficiently well-defined macroscopic state that
the entropy is essentially well defined, and that $\eta$ is
essentially well defined in that the characteristic smallest
time-scale for fluid behavior is much smaller than the
characteristic decay time. The heavy meson gas system considered
in Ref.~[3] {\it is} in this class and {\it does} violate the
bound. Thus, regardless of the validity of the issue raised in
Ref.~[1], Ref.~[3] at a minimum establishes the existence of a
theoretically consistent example that demonstrates a limitation of
the class of systems for which such a general bound can hold.

The critique in Ref.~[1] is not aimed at the validity of the heavy
meson gas of Ref.~[3] as a counterexample to the bound {\it per
se}, but rather at the expectation that the bound should apply to
systems in that class.  The key insight is that for the bound to
hold, it might not be sufficient for both $\eta$ and $s$ to be
essentially well defined on their own terms (hydrodynamic and
thermodynamic, respectively), but may require the system to live
long enough so  one can {\it simultaneously} measure both $\eta$
and $s$. Given the connection between $\eta$ and $s$ in the bound,
it is not unreasonable that its application be limited to systems
in which $\eta$ and $s$ are simultaneously well-defined. However,
this restriction does not invalidate a heavy meson system as a
counterexample to the bound.

Reference [1] argues that  in order to simultaneously measure
$\eta$ and $s$ for the heavy meson system of Ref.~[3], $\eta$ must
be measured over a thermodynamic length scale associated with the
size of the minimum system for which $s$ is well defined (which
scales as $\exp(\xi^4/3)$) rather than the hydrodynamic scale
(which is a power law in $\xi$). While $\eta$ {\it is} essentially
well defined on the hydrodynamic scale, it need not be over the
much larger thermodynamic scale,  since the metastable fluid
presumably decays before a measurement over this scale is
complete. Thus it is argued that the heavy meson system does not
have $\eta$ and $s$ simultaneously well defined.

However, it is {\it not} necessary to measure $\eta$ over the
thermodynamic length scale identified in [1] to have $\eta$ and
$s$ essentially well defined at the same time. The thermodynamic
limit is determined by a system's {\it volume} and not its length
scale (provided that all lengths of the system are large compared
to the thermal wavelength---as is the case for the heavy meson
system in Ref.~[3]).  Consider, for example, the stereotypical
setup for measuring the viscosity: fluid is contained between two
parallel rectangular plates whose cross-sectional size is $A$ and
whose separation is $d$ with $d \ll A^{1/2}$; $\eta$ is determined
by the force needed to keep one plate moving with fixed velocity
relative to the other. For the heavy meson system of Ref.~[3] this
setup means that the system is in the thermodynamic limit in the
sense that $s$ is essentially well defined if $A d \gg N_f \sim
\exp(\xi^4)$; $\eta$ is essentially well defined provided that $d$
is much larger than the mean-free path, $l$, ($l \sim \xi^4$)
while being much smaller than the scale characterizing decay
(which scales as a power law in $\xi$ times $l$ [4]).  Thus, if
$A=\alpha \exp(\xi^4)$ and $d=\beta \xi^4$ with $\alpha$ and
$\beta$ sufficiently large constants, the system at large $\xi$
will violate the proposed bound with $\eta$ and $s$ determined
simultaneously and each essentially well defined.

Numerous insightful comments of D.T.~Son, A.~Cherman and P.~Hohler
are gratefully acknowledged. This work was supported by the U.S.
Department of Energy.\vspace{.05in}

{\it \noindent Thomas D. Cohen\\
Department of Physics \\
University of Maryland\\College Park, MD 20742  }
\renewcommand{\labelenumi}{ {[}\theenumi{]} }
\begin{enumerate}
\item D.T.~Son, arXiv:0709.4651.
 \item P.~Kovtun, D.T.~Son and
 A.O.~Starinets, JHEP {\bf 0310} (2003) 064.
 \item T.D~Cohen,
Phys.~Rev.~Lett. {\bf 99} (2007) 021602.
 \item A.~Cherman,
T.D.~Cohen and P.M.~Hohler, arXiv:0708.4201.
\end{enumerate}
\end{document}